\documentclass[aps,prd,reprint,floatfix,superscriptaddress,lengthcheck,sort&compress,letterpaper,nofootinbib]{revtex4-1}

\usepackage{graphicx}
\usepackage{dcolumn}
\usepackage{bm}
\usepackage{dsfont}
\usepackage{amsmath}
\usepackage{amssymb}
\usepackage{microtype}
\usepackage[usenames]{color}
\usepackage{multirow}
\usepackage[colorlinks=true,linkcolor=blue,citecolor=blue,urlcolor=blue]{hyperref}

\newcommand{\gS}[1]{#1\!\!\!\!\!\not~}

\newcommand{\pslash}{\gS{p}}

\newcommand{\ii}{\textrm{i}}
\newcommand{\beq}{\begin{equation}}
	\newcommand{\eeq}{\end{equation}}
\newcommand{\bal}{\begin{align}}
	\newcommand{\eal}{\end{align}}

\begin{document}
	
	\title{The light scalars: four- vs. two-quark states in the complex energy plane from Bethe-Salpeter equations}
	
	\author{Nico Santowsky}
	\email[e-mail: ]{nico.santowsky@theo.physik.uni-giessen.de}
	\affiliation{Institut f\"ur Theoretische Physik, Justus-Liebig Universit\"at Gie{\ss}en, 35392 Gie{\ss}en, Germany}
	\affiliation{Helmholtz Forschungsakademie Hessen f\"ur FAIR (HFHF),
		GSI Helmholtzzentrum f\"ur Schwerionenforschung, Campus Gie{\ss}en, 35392 Gie{\ss}en, Germany}
	\author{Christian S. Fischer}
	\email[e-mail: ]{christian.fischer@theo.physik.uni-giessen.de}
	\affiliation{Institut f\"ur Theoretische Physik, Justus-Liebig Universit\"at Gie{\ss}en, 35392 Gie{\ss}en, Germany}
	\affiliation{Helmholtz Forschungsakademie Hessen f\"ur FAIR (HFHF),
		GSI Helmholtzzentrum f\"ur Schwerionenforschung, Campus Gie{\ss}en, 35392 Gie{\ss}en, Germany}

	\date{\today}
	
	\begin{abstract}
		
		We study the dynamical generation of scalar mesons in the light quark sector ($q\in\{u,d,s\}$) and calculate the masses 
		and widths for the $f_0(500),a_0(980)$ and $f_0(980)$. To this end we study the mixing of conventional $q\bar{q}$ and 
		`exotic' $q\bar{q}q\bar{q}$ states via a coupled set of two-body Bethe-Salpeter equations based on a symmetry-preserving 
		truncation of the underlying Dyson-Schwinger equations. This allows us to determine the dominant components of each state. 
		Furthermore, we extend our previous framework into the complex energy plane such that we can study the analytic structure
		of the states in question and extract their width.		
		At the physical point of small quark masses, the $\sigma$ meson is predominantly a $\pi\pi$ resonance. Consequently, its 
		mass and width is driven by the effects of chiral symmetry breaking. At larger quark masses, however, the conventional 
		$q\bar{q}$ components take over. Furthermore, we find a strong molecular $K\bar{K}$ component for both, the strange-light 
		$f_0(980)$ and the $a_0(980)$.
	\end{abstract}
	
	\maketitle
	
	
	\section{\label{sec:1}Introduction}
	There is a long history around the puzzling nature of the light scalar mesons. 
	Below 1~GeV, three of these states are experimentally observed: the very broad $f_0(500)$ (also known as $\sigma$, formerly $f_0(600)$), 
	the $f_0(980)$ and the $a_0(980)$. Originally considered as ordinary $q\bar{q}$ states, the mass ordering does not fit to naive quark 
	model predictions: according to the non-relativistic assignment of parity, ${P=(-1)^{L+1}}$, scalar mesons are p-waves and should 
	be heavier than their s-wave pseudoscalar and vector counterparts and are expected above $1\:\textrm{GeV}$. This is not the case 
	for the states in question. Furthermore, a conventional pure $q\bar{q}$ interpretation would expect the isospin partners
	$f_0(500)$ and $a_0(980)$ in a similar mass region in analogy to the $\rho/\omega$ mesons in the vector nonet. 
	A four-quark interpretation, however, would steer that mass ordering towards the observed one: The $f_0(500)$ would be the ground 
	state consisting of four light quarks, whereas the $a_0(980)$ and the $f_0(980)$ are isospin partners containing two light and 
	two strange quarks \cite{Jaffe:1976ig}. The small mass of the $f_0(500)$ is then explained by its s-wave nature and the fact 
	that it is essentially a $\pi\pi$ resonance \cite{Pelaez:2015qba}. This brings chiral effects into play: the pions as the pseudo 
	Goldstone bosons of QCD are considerably lighter than the constituent quarks appearing in quark model calculations. 
	
	The four-quark nature of the light scalar mesons also explains their decay pattern: the $f_0(500)$ with a strong $\pi\pi$ component
	immediately decays into two pions which results in a large decay width, whereas the $a_0/f_0(980)$ could establish a strong molecular
	$K\bar{K}$ component near the threshold \cite{Weinstein:1990gu}, entailing a narrower width. The physics of these systems has been
	explored in a variety of approaches such as lattice QCD \cite{Prelovsek:2010kg,Alexandrou:2012rm,Wakayama:2014gpa,Dudek:2016cru,Briceno:2016mjc,Briceno:2017qmb,Alexandrou:2017itd},
	dispersion theory \cite{GarciaMartin:2011jx,Londergan:2013dza,Pelaez:2017sit}, chiral effective field theory
	\cite{Pelaez:2003dy,Pelaez:2006nj,Hooft:2008we,RuizdeElvira:2010cs,Guo:2016zep,Ahmed:2020kmp}, 
	Dyson-Schwinger equations~\cite{Heupel:2012ua,Eichmann:2015cra,Eichmann:2020oqt} and model studies
	\cite{Giacosa:2006tf,Ebert:2008id,Parganlija:2010fz} supporting the predominant non-$q\bar{q}$-nature of these states. 
	
	It is then interesting to study the details of the internal structure of these states. In general, a four-quark structure accounts 
	for three different possibilities for internal configurations: (i) three- and four-body forces may be dominant generating a tightly 
	bound object without internal clustering and (ii) two-body forces may be dominant allowing for diquark-antidiquark
	clustering or (iii) meson-meson clusters. In the case of unequal quark masses the latter case may be split into (iiia) two heavy-light
	meson clusters or (iiib) one heavy and one light meson. These configurations may be superimposed and in addition there may be 
	(sizeable) admixtures from conventional $q\bar{q}$ configurations. Close-by thresholds such as $K\bar{K}$ may strongly affect these 
	mixtures as may be expected for the $a_0(980)$ and the $f_0(980)$.
	
	Although every state could potentially be modelled by only considering \textit{one} certain structure, a proper description 
	requires a treatment as an overlap of all the different components. This makes precise calculations of masses and decay widths 
	difficult and even affects the extraction of those quantities from experiment resulting in large uncertainties \cite{Zyla:2020zbs}.
	The PDG-estimate for the pole position of the $f_0(500)$ is given by ${400\dots550-\ii(200\dots350)\:\textrm{MeV}}$, for the $f_0(980)$ 
	the estimate reads ${(990\pm20)-\ii(5\dots50)\:\textrm{MeV}}$ and the $a_0(980)$ has values of ${(980\pm20)-\ii(25\dots50)\:\textrm{MeV}}$~\cite{Zyla:2020zbs}. 
	
	In this work we calculate the pole position of the $f_0(500)$, $f_0(980)$ and $a_0(980)$ in the complex energy plane from coupled Bethe-Salpeter equations (BSEs) that allow for meson-meson as well as diquark-antidiquark components in the four-quark components 
	and allow for mixing with conventional $q\bar{q}$ states \cite{Santowsky:2020pwd,Kvinikhidze:2021kzu}. We are therefore in a 
	position to determine the dominant part of the wave function of the different states. Previous approaches in this framework 
	have been restricted to extrapolations towards real and time-like momenta and were therefore not able to extract the nature 
	of the states \cite{Heupel:2012ua,Eichmann:2015cra}. Thus one could not distinguish between bound states, resonances or virtual 
	bound state. In this work we lift this restriction and provide access to the second Riemann-sheet by direct calculations in 
	the complex energy plane. We then investigate the significance of chiral effects by varying the quark mass systematically 
	and study the corresponding variation of dominant components. 
	
	The paper is structured as follows: In section \ref{sec:2} we discuss the coupled system of BSEs. In section \ref{sec:3} we then
	explain technical details, in particular the methods used to explore the complex energy plane. We then display our numerical 
	results for the masses and widths of the states in section \ref{sec:4} and conclude with a number of remarks.
	\section{\label{sec:2}Four- and two-body Bethe-Salpeter equations}
	\noindent
	This section is a brief summary of the construction of a coupled system of BSEs for the four- and two-quark systems. More
	details can be found in \cite{Santowsky:2020pwd}. An alternative derivation is presented in \cite{Kvinikhidze:2021kzu}.\vspace{4pt}\\
	\textbf{The four-body equation.~} We derive the equations of motion for $n$-quark bound states, the Bethe-Salpeter equations, from the ($2n$)-quark scattering matrix $T^{n}$ and the corresponding scattering kernel $K^{(n)}$:
	\begin{equation}
		T^{(n)} = K^{(n)}+K^{(n)}G_0^{(n)}T^{(n)}
		\label{eq-dysonsum}
	\end{equation}
	Since an $n$-quark state induces a singularity at the bound state total momentum $P^2$, one applies a pole ansatz thereby defining 
	the Bethe-Salpeter amplitudes $\Gamma^{(n)}$ as the residue on the mass-shell:
	\begin{equation}
		T^{(n)}\xrightarrow{P^2\rightarrow-M^2}\frac{\Gamma^{(n)}\bar\Gamma^{(n)}}{P^2+M^2} \label{eq-bsaansatz}
	\end{equation}
	These amplitudes carry the full Dirac, colour and flavour structure of the state in question. Whereas bound states appear as poles 
	on the timelike, negative real momentum axis, resonances are seen in the second Riemann sheet of the complex plane according to
	${M\rightarrow (M+\ii\:\Gamma/2)}$, where $\Gamma$ is the decay width of the state.
	
	The full $4$-quark BSE with the scattering kernel $K^{(4)}$ reads
	\begin{align}
		\Gamma^{(4)}& =K^{(4)}G_0^{(4)}\Gamma^{(4)}\\
		K^{(4)}& = \tilde{K}^{(2)}+\tilde{K}^{(3)}+\tilde{K}^{(4)}
	\end{align}
	with $\tilde{K}^{(2,3,4)}$ containing two-, three- and four-quark irreducible contributions. In the following we assume that 
	internal two-body interactions dominate and neglect the latter two setting $\tilde{K}^{(3)}=\tilde{K}^{(4)}=0$. As discussed 
	in the introduction, this restricts our framework to the description of states with dominating internal meson-meson and/or 
	diquark-antidiquark configurations which may be expected for the light scalar nonet. 
	
	The remaining two-quark scattering kernel can be decomposed into direct interactions between two quarks:
	\begin{align}
		\tilde{K}^{(2)}	&	=\underbrace{{K}^{(2)}_{12}S^{-1}_3S^{-1}_4+{K}^{(2)}_{34}S^{-1}_1S^{-1}_2
			-{K}^{(2)}_{12}{K}^{(2)}_{34}}_{=:\tilde{K}^{(2)}_{(12)(34)}}
		+ \text{perm.}\nonumber\\
		&   = \sum_a \tilde{K}^{(2)}_a \label{eq-2bodykernels}
	\end{align}
	\textbf{The two-body approximation.~} So far we still need to solve a four-body problem. This has been performed  
	in \cite{Eichmann:2015cra,Wallbott:2019dng,Wallbott:2020jzh}. In order to be able to take into account the additional 
	mixing with $q\bar{q}$-state, we further simplify the equation along the lines of \cite{Heupel:2012ua,Santowsky:2020pwd} 
	into a two-body problem with effective meson and diquark degrees of freedom. To this end we define a four-body $T$ matrix 
	$T_a$ that is generated by $\tilde{K}_a^{(2)}$,
	\begin{equation}
		T_a = \tilde{K}^{(2)}_a + \tilde{K}^{(2)}_a G_0^{(4)} T_a = \tilde{K}^{(2)}_a +  T_a G_0^{(4)} \tilde{K}^{(2)}_a\,, \label{Ta}
	\end{equation}
	which splits the four-body BSA $\Psi$ into three separate parts:
	\begin{align}
		\Psi = \sum_a \tilde{K}^{(2)}_a G_0^{(4)}\:\Psi := \sum_a \Psi_a\,.
	\end{align}
	Acting with $T_a G_0^{(4)}$ onto $\Psi$ and using (\ref{Ta}) one then obtains
	\begin{equation}
		\Psi_a=T_a \,G_0^{(4)}\,(\Psi-\Psi_a)=\sum_{b\neq a}\,T_a\,G_0^{(4)}\,\Psi_b\,, \label{eq-masterbse}
	\end{equation}
	which is still an exact four-body equation apart from neglecting the kernels $\tilde{K}^{(3)}$ and $\tilde{K}^{(4)}$.
	The two-body approximation of this four-body equation then amounts to replacing $T_a$ with a pole ansatz analogously 
	to Eq.~(\ref{eq-bsaansatz}). We thus obtain
	\begin{equation}
		\Psi_a = \left(\Gamma_{12}\otimes\Gamma_{34}\right)\:G_0^{(2,2)}\:\Phi_{a} \label{eq-4bsastructure}
	\end{equation}
	for $a=(12)(34)$ and similar expressions for the other combinations. Here $G_0^{(2,2)}$ is a combination of two meson 
	propagators or a diquark and an antidiquark propagator, respectively, and $\Gamma_{ij}$ are the corresponding two-body 
	Bethe-Salpeter amplitudes. The representation Eq.~(\ref{eq-4bsastructure}) is in some sense a `physical basis' in that 
	it builds a representation of $\Psi_a$ in terms of reduced internal Dirac, flavor and color structure from a physical 
	picture. The algebraic structure of the tetraquark-meson and tetraquark-diquark vertices $\Phi_a$ depend on the 
	respective quantum numbers of the investigated four-quark state. For scalar four-quark states and (pseudo)scalar 
	ingredients, e.g., those amplitudes are flavor and color singlets and Lorentz scalars.
	
	With Eq.~(\ref{eq-4bsastructure}), we effectively solve for the vertices $\Phi_a$ while making use of solutions of the two-quark
	BSE for the amplitudes $\Gamma_{ij}$. The interaction kernel elements for the internal vertices $\Phi_a$ are quark exchange diagrams.\vspace{4pt}\\ \\
	\textbf{Mixing with conventional $\boldsymbol{q\bar{q}}$ states.~}
	We extend the truncation for the four-body equation (\ref{eq-masterbse}) by adding a phenomenologically motivated two-quark component
	into the Bethe-Salpeter amplitude. For $a=(12)(34)$ in Eq.~(\ref{eq-4bsastructure}), this amounts to
	\begin{equation}
		\Psi_a = \left(\Gamma_{12}\otimes\Gamma_{34}\right)\:G_0^{(2,2)}\:\Phi_{a} +
		K^{(2)}_{13} K^{(2)}_{24} S_{34} G_0^{(2)}\Gamma^*_{12}  +\text{perm.}
		\label{eq-4bsastructure_v2}
	\end{equation}
	where $\Gamma^*_{12}$ is the Bethe-Salpeter amplitude of a quarkonium state with the same quantum numbers as the four-body state connected
	to the propagator lines $1$ and $2$, and the quark propagator $S_{34}$ connects the lines $3$ and $4$.
	This is equivalent to extending the physical basis discussed above with another possible basis element.
	Note that these extensions only
	appear in those $\Psi_a$ with two-body interactions between quark-antiquark pairs, i.e. meson-meson contributions. They do not appear in
	the $\Psi_a$ with two-body interactions between quark-quark pairs, i.e. in diquark/antidiquark contributions.
	
\begin{figure}
		\includegraphics[width=8.7cm]{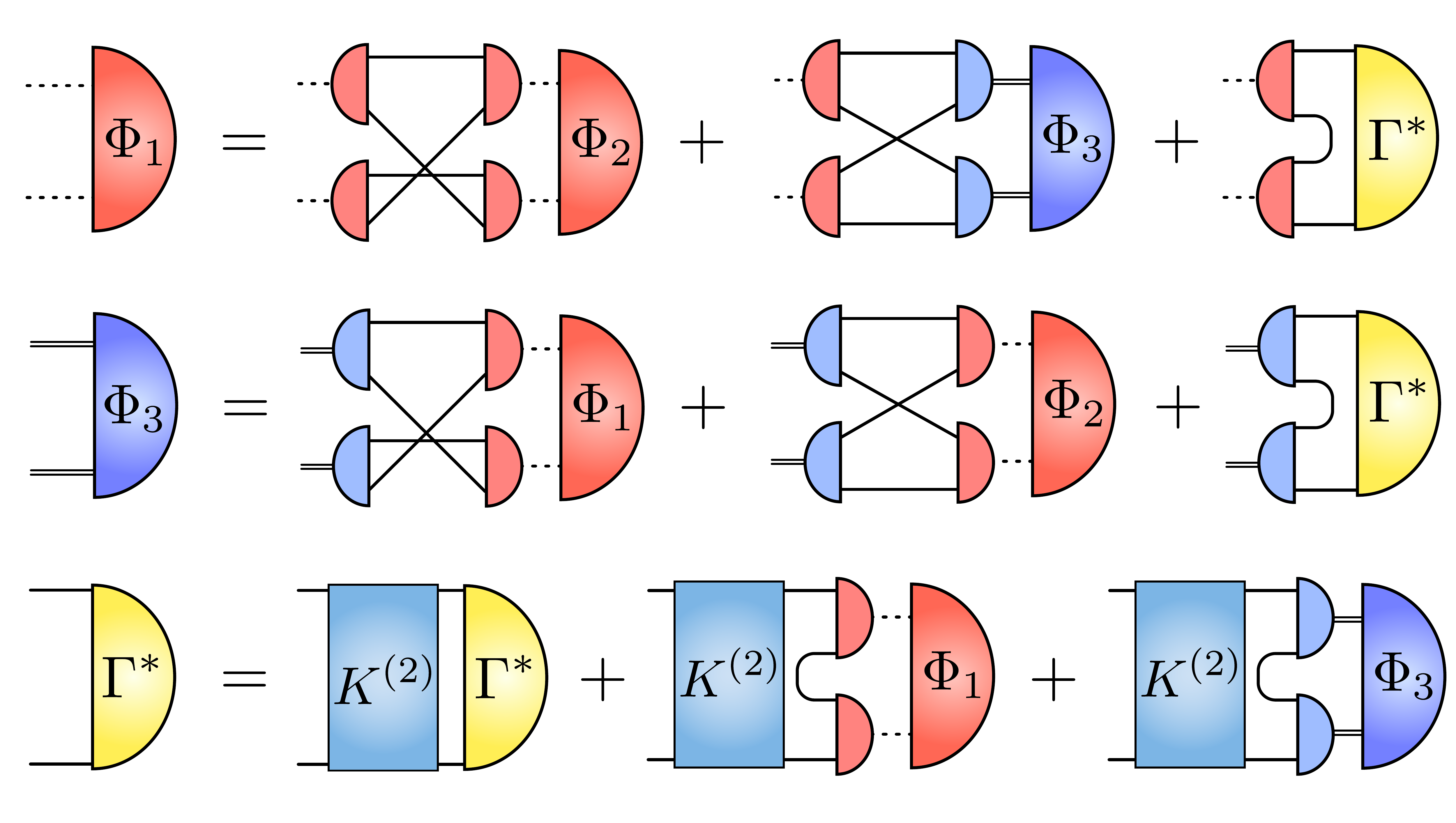}
		\caption{The coupled system of BSEs for one of the meson-meson contributions (red; first line) and the
			diquark/antidiquark contribution (blue; second line) to the four-quark state as well as the
			coupled BSE for the quark-antiquark state (yellow; third line) with the same quantum numbers.
			Not shown are the additional BSE for the second meson-meson contribution $\phi_2$ (which is redundant in our case and only contributes if we use four equal quark masses)
			and another diagram for $\Gamma^*$ containing $\phi_2$ instead of $\phi_1$.   \label{fig-diagram2}}
	\end{figure}

	The corresponding two-quark $T$-matrix that contains this quarkonium state is determined by (cf. Eq.~(\ref{eq-dysonsum}) with $n=2$),
	\begin{equation}
		T^{(2)}=\left(\mathds{1}-K^{(2)}G_0^{(2)}\right)^{-1}K^{(2)}\,,
	\end{equation}
	and the four-quark $T$-matrix that contains the two-body component added in
	Eq.~(\ref{eq-4bsastructure_v2}) is denoted by
	\begin{equation}
		T_a^{(4,2)} = K^{(2)}_{13} K^{(2)}_{24} S_{34} G_0^{(2)} T_{12}^{(2)} G_0^{(2)} S_{34} K^{(2)}_{24} K^{(2)}_{13}.
	\end{equation}
	As a result, the master equation (\ref{eq-masterbse}) then contains the two-body equation for $\Gamma^*$ as an additional element and the equations
	for the four-body meson-meson and diquark-antidiquark components of the full BSA are modified by additional terms containing $\Gamma^*$. The resulting
	system of equations is shown diagrammatically in Fig.~\ref{fig-diagram2}.
	We observe $q\bar{q}$ contributions in both BSEs for the meson-meson
	and  diquark-antidiquark components of the four-body amplitudes as well as a back-coupling of the meson-meson and diquark-antidiquark components into the
	$q\bar{q}$ equation.
	
	Note that there has been some debate about the structure of the mixing term in the kernel for the conventional meson state
	\cite{Blankleider:2021odu,Santowsky:2021lyc}. Numerically, it turns out that both formulations,
	Refs.~\cite{Santowsky:2020pwd,Kvinikhidze:2021kzu} are almost equivalent and are equally well suited for the qualitative 
	discussion presented in this work. While the formulation of Kvinikhidze and Blankleider \cite{Kvinikhidze:2021kzu} is 
	probably more elegant and certainly more efficient in terms of CPU-time, we still stick to the formulation presented 
	in \cite{Santowsky:2020pwd} in order to ensure direct comparability with our previous work.
	
	\section{\label{sec:3} Technical details}
	
	\subsection{\label{sec:3.1} Quark propagator, mesons and diquarks}
	
	The elements necessary to solve the coupled system of BSEs in Fig.~\ref{fig-diagram2} are dressed quark propagators,
	meson and diquark Bethe-Salpeter amplitudes and the corresponding propagators. All these ingredients are determined 
	consistently from an underlying quark-gluon interaction that respects chiral
	symmetry. The technical details of these types of calculations have been described in many works, see e.g. \cite{Maris:2003vk,Eichmann:2016yit,Sanchis-Alepuz:2017jjd} for reviews, thus we only give a short summary here.
	
	The exact Dyson-Schwinger equation for the dressed quark propagator reads
	\begin{equation}\label{eq-quarkdse}
		S^{-1}_{\alpha\beta}(p) = Z_2 \left( i\pslash + Z_mm_0 \right)_{\alpha\beta}
		+ C_F \int_q \mathcal{K}_{\alpha\alpha'\beta'\beta} \,S_{\alpha'\beta'}(q)\,,
	\end{equation}
	with wave-function renormalization constant $Z_2$, bare quark mass $m_0$ and the Casimir $C_F=4/3$ for $N_c=3$ from the color trace.
	The interaction kernel $\mathcal{K}_{\alpha\alpha'\beta'\beta}$ contains the dressed gluon propagator as well as
	one bare and one dressed quark-gluon vertex. The Greek subscripts refer to color, flavor and Dirac structure. In previous treatments of the four-quark problem \cite{Heupel:2012ua,Eichmann:2015cra,Wallbott:2019dng,Wallbott:2020jzh,Santowsky:2020pwd}, the rainbow-ladder
	approximation has been used and we adopt the same interaction here. Then the kernel can be written as
	\begin{equation}\label{eq-RLkernel}
		\mathcal{K}_{\alpha\alpha'\beta\beta'} =  Z_2^2 \, \frac{ 4\pi \alpha(k^2)}{k^2} \,
		T^{\mu\nu}_k \gamma^\mu_{\alpha\alpha'} \,\gamma^\nu_{\beta\beta'},
	\end{equation}
	with the transverse projector $T^{\mu\nu}_k=\delta^{\mu\nu} - k^\mu k^\nu/k^2$. In this formulation, both the gluon dressing
	function and the vector part $\sim \gamma^\mu$ of the quark-gluon vertex have been absorbed into an effective running
	coupling $\alpha(k^2)$ which is taken from Ref.~\cite{Maris:1999nt} and has been discussed in detail e.g. in \cite{Eichmann:2016yit}.
	The explicit expression is given in appendix \ref{app:model}.
	This truncation guarantees the correct logarithmic behaviour of the quark at large momenta. Most importantly in the present 
	context, it also allows for the preservation of the axialvector Ward-Takahashi identity by using the same interaction kernel
	in the Bethe-Salpeter equations for the mesons and diquarks.
	
	With the quark propagator from Eq.~(\ref{eq-quarkdse}) and the quark-(anti-)quark interaction kernel Eq.~(\ref{eq-RLkernel}) we then solve the
	Bethe-Salpeter equations for light pseudoscalar mesons and scalar diquarks, which are the leading components (in terms of smallest masses)
	of the two-body composition of our scalar four-quark state. The explicit representation of the BSA in terms of (four) Dirac, flavor and color
	components as well as details on the technical treatment of meson BSEs can be found in the review articles \cite{Eichmann:2016yit,Sanchis-Alepuz:2017jjd}.
	The meson/diquark propagators are then calculated via $T=\Gamma D \bar{\Gamma}$ and Eq.~(\ref{eq-dysonsum}). 
	The results for the most important light-quark states are given in Tab.~\ref{tab:mesons}. The systematic error given in the table
	(and also included in all subsequent results) is determined from model variations, see appendix \ref{app:model} for details.
	\begin{table}[t]
		\normalsize
		\begin{tabular}{|c|c|c|c|c|}
			\hline
			~		&	$m_\pi$	&	$m_K$		&	$m_{qq,0^+}$	&	$m_{sq,0^+}$ \\\hline
			[GeV]	&	0.139(2)&	0.500(2)	&	0.801(31)		&	1.108(57)	\\\hline
		\end{tabular}
		\caption{Masses of selected pseudoscalar meson and scalar diquark states. The error stems from model variations;
			details are given in appendix \ref{app:model}. \label{tab:mesons}}
	\end{table}
	\begin{figure}[t]
		\includegraphics[width=7.5cm]{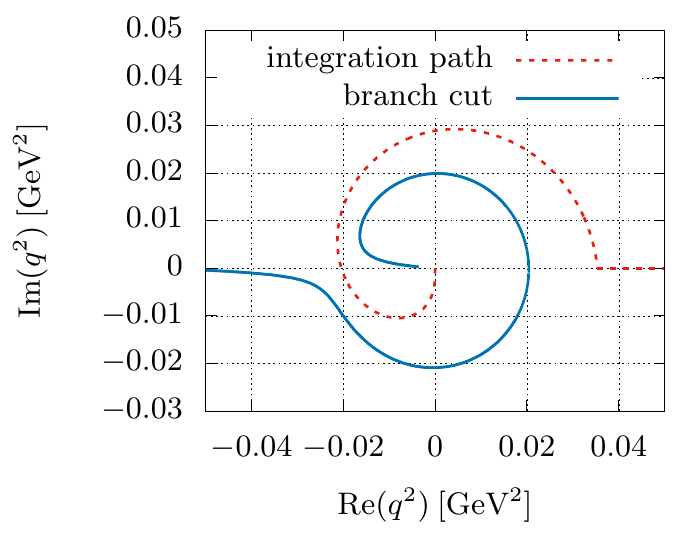}
		\caption{The branch cut in the radial integration domain of the two-body BSE (blue, solid line) and the chosen integration path that circumvents the cut (red, dashed) above the two-meson threshold. In this example, the cut stems from the $\pi\pi$ threshold with an input mass and width of ${(M,\Gamma)=(0.40,0.01)\:\textrm{GeV}}$ for the composite state.}\label{fig-pathdef}
	\end{figure}
	
	Since the pseudoscalar meson and scalar diquark states are dominated by their leading Dirac structure, we only take those
	into account in BSAs appearing internally in diagrams. For the external BSA in the two-body equation of the scalar meson
	we use the full structure, i.e. all four amplitudes.
	
	With this input, we solve the eigenvalue problem of the two-body BSE by introducing a general, momentum-dependent eigenvalue $\lambda(P^2)$ and search for the total
	momentum $P^2$ that corresponds to $\lambda=1$. 
	~
	
	\subsection{\label{sec:3.2} Threshold effects}\label{complex}
	A technical difficulty is induced by the presence of poles in the equation, e.g. the ones occurring in the meson and diquark propagators.
	They restrict a calculation of the eigenvalue curve on the real momentum axis to values below a critical mass $M_\textrm{crit}=2m$, 
	where $m$ is the mass of the lightest meson or diquark. Above this critical value, a branch cut opens that separates two Riemann 
	sheets. The first one is directly accessible using techniques of path deformation that have been developed and explored in the context
	of conventional mesons in \cite{Weil:2017knt,Williams:2018adr,Miramontes:2018omq,Miramontes:2019mco,Santowsky:2020pwd}. 
	In this work, for the first time, we apply this technique to the four-quark-case using a path deformation in the radial integral 
	that is displayed in Fig.~\ref{fig-pathdef}. This enables direct calculations on the first Riemann sheet. We subsequently continue 
	into the second sheet analytically. The corresponding technical details are provided in appendix \ref{app:extra}.

	\section{\label{sec:4} Results}
	
	\subsection{The ${\sigma/f_0(500)}$ candidate as a ${q\bar{q}(q\bar{q})}$ state}\label{sec:4.1}
	
	\begin{figure}[t]
		\includegraphics[width=8cm]{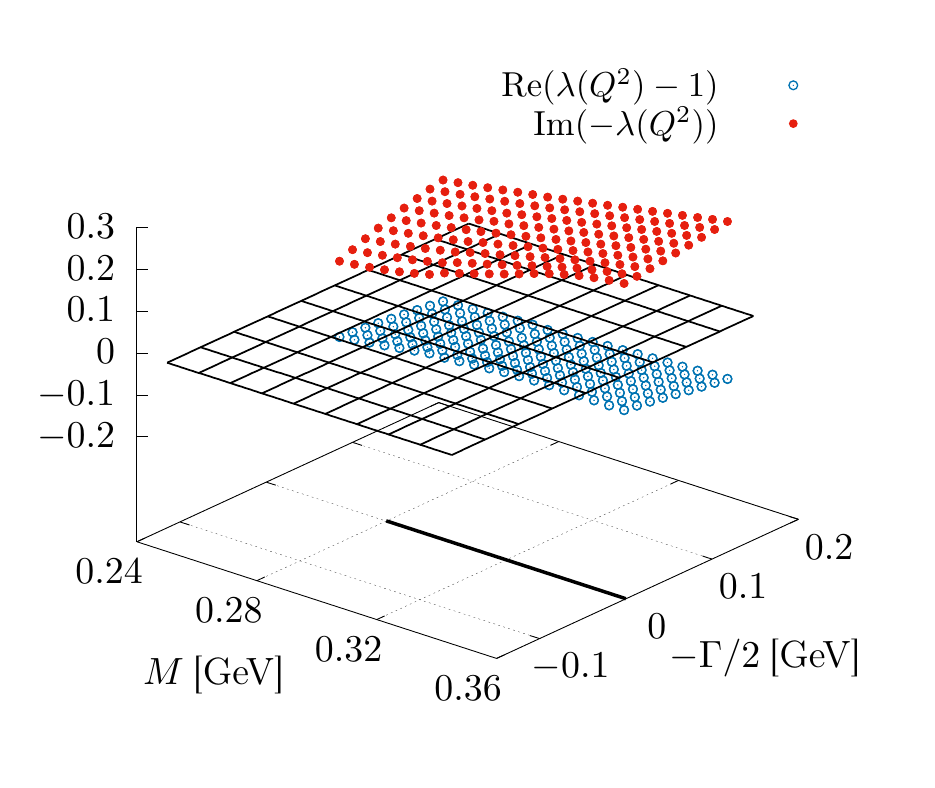}
		\caption{The eigenvalue curve of the fully mixed $\sigma/f_0(500)$ candidate in the complex plane. Red, filled dots: The real part subtracted by 1. Blue, unfilled dots: The (negative) imaginary part. A solution of the equation is found if both curves match each other at a value of zero (visualized as a black net). On the bottom, the domain of the branch cut is shown as a thick, black bar.}\label{fig-evcurve}
	\end{figure}
	
	We first focus on the isoscalar scalar state with four light quarks, i.e. for a candidate of the $\sigma/f_0(500)$. Its 
	quantum numbers allow for internal structures in terms of two pseudoscalar mesons, a scalar diquark/antidiquark pair and 
	the admixture of a conventional scalar $q\bar{q}$-component. In symbolic notation, this reads:
	\begin{equation}
		f_0(500) = \pi\pi + 0^+0^+ + q\bar{q}
	\end{equation}
	In order to identify the most important components, we switch individual components on and off thus distinguishing between 
	the following setups:
	\begin{itemize}
		\item $\pi\pi+0^+0^++q\bar{q}$ (`fully mixed')
		\item $\pi\pi+0^+0^+$
		\item $\pi\pi$
		\item $q\bar{q}$ (pure 2-quark)
	\end{itemize}
	In Fig.~\ref{fig-evcurve} we see the eigenvalue curve of the `fully mixed' state in the ($M,\Gamma$) plane.
	
	Given that the imaginary part is antisymmetric under mirroring on the real axis within one Riemann sheet, we see that a branch 
	cut opens for the imaginary part of the curve above the two-pion threshold at $(M,\Gamma)=(2m_\pi,0)\approx(0.278,0)$ MeV. 
	We extrapolate into the second Riemann sheet, i.e. $\Gamma/2>0$, with techniques described above and extract solutions of the BSE
	at points where ${\textrm{Re}(\lambda-1)=\textrm{Im}(\lambda)=0}$ holds. The solutions for the respective setups are given 
	in Tab.~\ref{tab-f0500}.

	We see that all results with $\pi\pi$ contributions have masses around 300\:MeV. In particular, adding diquarks 
	to the $\pi\pi$ component induces hardly any changes. This confirms the results of Refs.\cite{Heupel:2012ua,Eichmann:2015cra}:
	diquarks are almost irrelevant for the internal structure of the $\sigma/f_0(500)$. 
	Furthermore, the additional $q\bar{q}$ component is just lowering the mass and the width slightly. 
	In comparison, the pure $q\bar{q}$ state is located far above the ones with four-quark contributions 
	(661 vs. 291~MeV). Again this confirms previous findings \cite{Santowsky:2020pwd}.
	Comparing with experiment, one would expect the mass of the conventional $q\bar{q}$-state to be 
	much heavier, i.e. far above 1 GeV. This is a well-known artefact of the rainbow-ladder truncation 
	that we employ in the two-body quark-antiquark interaction: while the masses of pseudoscalar and 
	vector mesons are obtained in the correct mass range, the masses of scalars and axialvectors are 
	notoriously much too small. This is remedied only in beyond rainbow-ladder calculations, which indeed 
	push the mass of the conventional scalar $q\bar{q}$ state well above 1 GeV \cite{Chang:2011ei,Williams:2015cvx}.
	We come back to this point in section \ref{sec:4.4}. In any case, heavier conventional states 
	will make the mixing effects even smaller than observed. We therefore arrive at the robust conclusion 
	that the $\sigma/f_0(500)$ state is strongly dominated by the $\pi\pi$ four-quark component with almost 
	no contributions by diquark-antidiquark components and only few-percent corrections
	due to mixing with quarkonium components. 
	
	Compared to the mass and width of the $f_0(500)$ extracted from pion scattering experiments
	with dispersive tools, $M-\ii\:\Gamma/2=449^{+22}_{-16}-\ii(275\pm12)$ MeV \cite{Pelaez:2015qba}, 
	we find a smaller mass by about $30\%$. Also, the width is too small. Clearly, our model
	only retains qualitative aspects of the full quark-antiquark four-body equation, but not all quantitative
	aspects: the residual quark-exchange interaction between the meson and diquark components
	binds the four-quark state too strongly. Playing with the interaction strength, we find 
	that the pole location is especially sensitive to the $\pi\pi-\pi\pi$ and $\pi\pi-q\bar{q}$ 
	diagrams: artificially lowering those couplings e.g. by a factor of 0.7 results in almost
	double the width and we obtain $(M,\Gamma/2)=\big(309(14),232(54)\big)\:$MeV for the fully 
	mixed state. However, one should probably not attribute too much meaning to such artificial changes.
	In the end, it may very well be that quantitative results can only be obtained in the full
	four-body framework. This is subject of ongoing research. 
	Nevertheless we would like to stress that the two-body framework employed here is extremely
	valuable to gain qualitative insights first.

	\begin{table}
		\begin{normalsize}
			\begin{tabular}{ccc}
				\hline
				state						& $M$ [MeV]	& $\Gamma/2$ [MeV] 	\\\hline\hline
				$\pi\pi+0^+0^++q\bar{q}$	& 291(5)	& 121(22)			\\
				$\pi\pi+0^+0^+$				& 302(7)	& 148(31)			\\
				$\pi\pi$					& 301(7)	& 158(29)			\\
				$q\bar{q}$					& 661(8)	& 0					\\\hline
			\end{tabular}
		\end{normalsize}
		\caption{The masses and (half) widths of the different setups to describe the $\sigma/f_0(500)$ candidate. Errors stem from extrapolations and model variations, see the appendix for details.}\label{tab-f0500}
	\end{table}

	\subsection{Effects of dynamical chiral symmetry breaking: 
		The mixed $\boldsymbol{q\bar{q}(q\bar{q})}$ state with $\boldsymbol{q=u/d\rightarrow s}$
	}\label{sec:4.2}
	
	\begin{table}
		\begin{normalsize}
			\begin{tabular}{c|c||cccc}
				\hline
				\multicolumn{2}{c||}{$m_q$}            				& 3.8     & 20.0    & 50.0    & 85.5     	\\ \hline 
				\multicolumn{2}{c||}{$M_q(0)$}            			& 482     & 530     & 601     & 671     	\\ \hline 
				\multicolumn{2}{c||}{$q\bar{q}_{\small \textrm{PS}}$}        & 139     & 328     & 526     & 701     	\\ \hline \hline
				\multirow{2}{*}{$q\bar{q}$} 		& $M$			& 661(8)  & 739(6)  & 881(13) & 1073(10) 	\\
				& $\Gamma/2$ 	& 0       & 0       & 0       & 0       	\\ \hline
				\multirow{2}{*}{$q\bar{q}q\bar{q}$} & $M$  			& 302(7)  & 665(10) & 1045(9) & 1414(21) 	\\
				& $\Gamma/2$ 	& 148(31) & 110(41) & 74(22)  & 34(20)   	\\ \hline
				\multirow{2}{*}{mixed}   			& $M$        	& 291(5)  & 669(3)  & 827(3)  & 1047(12) 	\\
				& $\Gamma/2$ 	& 121(22) & 11(13)  & 0       & 0        	\\ \hline
			\end{tabular}
		\end{normalsize}
		\caption{The masses and (half) widths of the different setups to describe a $0(0^+)$ mixed $q\bar{q}(q\bar{q})$ state for varying 
			(current) quark masses $m_q$. We also give values for the resulting quark mass function $M(0)$ at $p^2=0$ and the masses of the
			pseudoscalar $q\bar{q}_{\small \textrm{PS}}$ which may appear as constituent in the scalar four/two-quark states. 
			All numerical values are given in MeV.}\label{tab-heavyheavy}
	\end{table}
	\begin{figure*}[t]
		\includegraphics[width=0.9\textwidth]{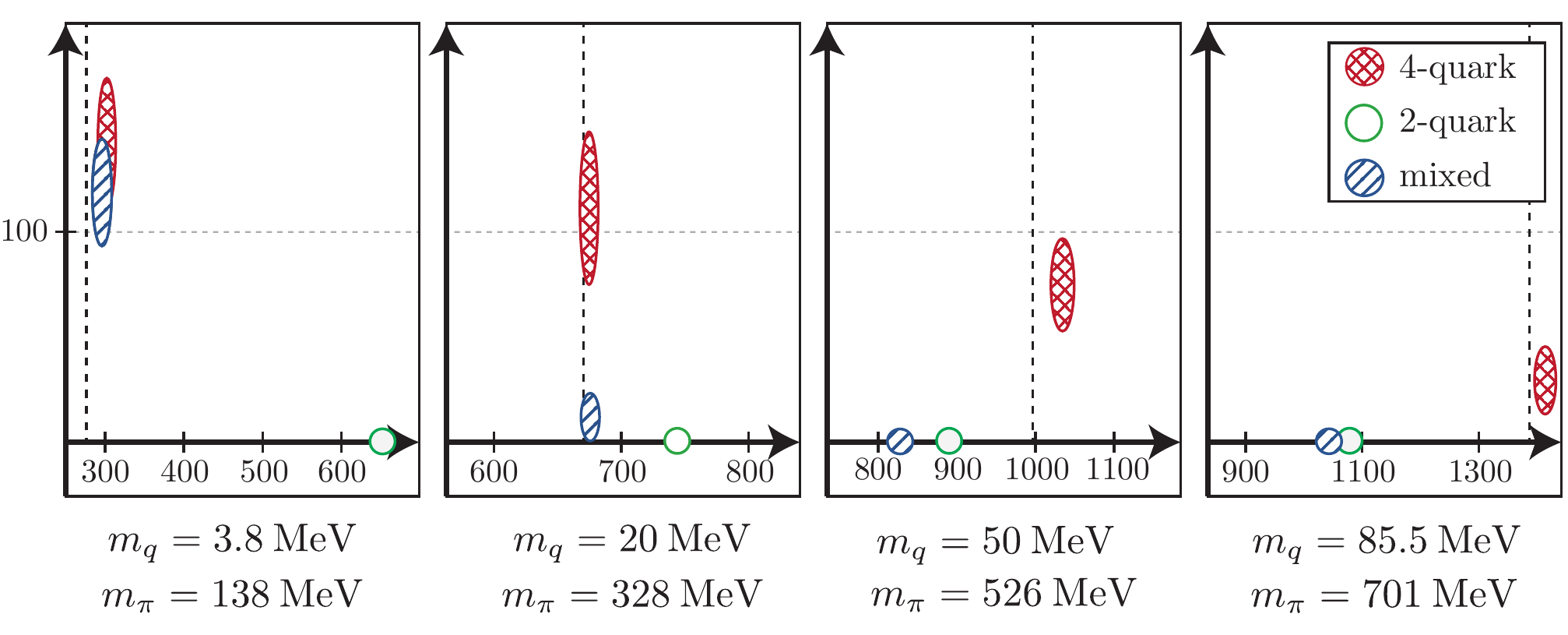}
		\caption{A visualization of the pole positions $(M,\Gamma/2)$ of a mixed $q\bar{q}(q\bar{q})$ state as we vary $q=u/d\rightarrow{}s$ with $m_u=m_d=3.8\:$MeV and $m_s=85.5\:$MeV by using coloured ovals. The sizes of the ovals roughly denote the errors. 
			Dashed lines denote the two-meson mass thresholds. All numerical values are given in MeV.}\label{fig-heavyheavy}
	\end{figure*}
	In order to highlight the crucial aspect played by dynamical chiral symmetry breaking in the internal structure of the $\sigma/f_0(500)$,
	we now increase the masses of the light quarks step by step until they reach values typical for the strange quark mass, which has been 
	fixed previously from the BSE for a $q\bar{s}$-state via the physical kaon mass. At each step we study the mixing of the four-quark
	state with the corresponding conventional meson state. The values for the masses and half widths of four equal-mass $q\bar{q}q\bar{q}$ 
	states with quark masses at the two end-points (physical light and strange quark masses) and two states in between are given in
	Tab.~\ref{tab-heavyheavy}. The results are visualized in Fig.~\ref{fig-heavyheavy}. We distinguish between pure 4-quark states
	($\pi\pi+0^+0^+$), pure 2-quark states ($q\bar{q}$) and mixed states. The results for the lightest quark mass, $m=3.8\:$MeV, equal 
	the ones we gave already in section~\ref{sec:4.1} for the $\sigma/f_0(500)$. Here, the state is dominated by the internal structure with
	the lightest components. Because of dynamical chiral symmetry breaking this is not the $q\bar{q}$-component with quark masses of about
	$M(p^2=0) \approx 400\dots500$ MeV, but the component with the two (pseudo-) Goldstone bosons, $\pi\pi$. When increasing the light quark 
	mass to 20 and 50~MeV, we observe that the clear four-quark dominance gets lost and we have significant mixing effects between the two- 
	and the four-quark components. The width decreases strongly and from $m_q=50\:\text{MeV}$ on, the mixed state is bound, i.e. the width is zero. 
	When arriving at the physical strange quark mass, we observe a state which is dominated by the two-quark component. Now the quark
	masses in the $s\bar{s}$ components are much smaller than the corresponding masses of the pseudoscalar $\pi\pi$-components (with
	strange quarks instead of light ones) and consequently it is energetically favourable to form a conventional meson instead of a 
	meson-meson or diquark/antidiquark state in our model. A more realistic description of the scalar $q\bar{q}$ state using a more
	advanced model or an inclusion of beyond RL contributions would probably suppress the importance of the $q\bar{q}$ component more
	due to the higher mass of the conventional meson.
	
	We believe this mass dependence of the composition of states is very instructive and may be important
	to keep in mind when interpreting lattice results at non-physical quark masses.
	
	\subsection{\label{sec:4.3} The 
		$\boldsymbol{f_0(980)}$ and $\boldsymbol{a_0(980)}$ 
		as mixed states}
	We now focus on the two lightest experimentally confirmed scalar states with strange quark content:
	the isoscalar $f_0(980)$ and the isovector $a_0(980)$. In the four-quark picture these are expected 
	to be $s\bar{s}q\bar{q}$ states with $q\in\{u,d\}$. Here, we consider the following decomposition 
	of the states with substructures motivated by leading decay channels:
	\begin{align}
		f_0(980) &= \underline{K\bar{K}} + \pi\pi + \underline{0^+0^+} + \underline{s\bar{s}} + \underline{(u\bar{u}+d\bar{d})} \notag\\
		a_0(980) &= \underline{K\bar{K}} + \pi\eta + \underline{0^+0^+} + \underline{(u\bar{u}-d\bar{d})} \label{eq-a0_f0_components}
	\end{align}
	In our calculations, we will only consider the underlined clusters for the following, technical reasons: 
	The $\pi\pi$ component in the $f_0$ does not fit to the valence quark content we assume. Its inclusion
	would require the (Zweig-suppressed) internal conversion of a $s\bar{s}$-pair into a light quark pair.
	We expect such a component only to contribute significantly to the width of the $f_0$ but not to its 
	mass. Furthermore, both, the $\pi\pi$ component in the $f_0$ and the $\pi\eta$ component for the $a_0$
	are technically extremely difficult to realise: either one includes residues due to the pion poles
	in the integration domain of the BSE explicitly, or one has to perform extrapolations to an extent 
	that makes even qualitative results questionable. We therefore defer the inclusion of both contributions
	to future work. We furthermore ignore all contributions of light quark-antiquark states for the moment,
	because, as discussed above, the conventional rainbow-ladder scalar states are much too low in mass and will 
	therefore flaw the calculation. Instead, the rainbow-ladder $s\bar{s}$ with a mass above 1 GeV serves 
	as a test state for our first exploration of mixing effects in this section. We will then
	refine our framework further by including the light quark contributions in section~\ref{sec:4.4}. 
	
	The results for different setups\footnote{Note that a calculation for a pure molecular $K\bar{K}$ 
		state or a pure diquark state is not possible in the present framework due to the off-diagonal 
		structure of the two-body 
		BSE (\ref{eq-masterbse}).} are shown in Tab.~\ref{tab-a0_and_f0}. We denote possible $f_0(980)$ 
	or $a_0(980)$ candidates by bullets ($\bullet$) in the leading columns.\footnote{Since we are 
		working in the isospin symmetric limit, the $a_0$ and the $f_0$ are mass-degenerate; therefore, 
		solution of the BSE corresponds to both. Differences in the flavour part of their respective 
		Bethe-Salpeter amplitudes do not affect the results.} 
	
	Let us first discuss the result for the four-quark state ($K\bar{K}+0^+0^+$) without admixures 
	from the conventional $s\bar{s}$-state. The pole location on the second Riemann-sheet is found 
	at $M-\ii\:\Gamma/2=1001(4)-\ii\:24(16)\:$MeV in a region which is expected from model-independent, 
	dispersive analyses: $(996\pm7)-\ii(25^{+10}_{-6})$ \cite{GarciaMartin:2011jx}. The admixture of 
	the $s\bar{s}$ component has a significant effect (1001 vs. 915\:MeV) and is probably too strong
	due to the proximity of the mass of the pure $s\bar{s}$-state (1073~MeV) to the unmixed four-body
	state. As discussed above, this proximity is an artefact. We will address this issue in 
	section~\ref{sec:4.4}. Nevertheless, the coupling of the $s\bar{s}$-state serves to address the 
	diquark-issue: Comparing same setups with and without diquarks reveal their contribution again
	to be only at the percent level (915 vs. 927~MeV). Non-zero widths for those two states (see the 
	lower two rows) could be considered as numerical artefacts as states below the decay threshold could
	not have a width different from zero.
	
	We thus conclude that our candidates for the $a_0$ and the $f_0$ both have a strong molecular $K\bar{K}$ 
	and a potentially non-negligible $q\bar{q}$ component, which is investigated further in the next section.

	\begin{table}[]
		\begin{normalsize}
			\begin{tabular}{ccccc}
				\hline
				$f_0$		& $a_0$		& setup                      & $M$      & $\Gamma/2$ \\ \hline \hline
				$\bullet$  	& $\bullet$	& $K\bar{K}+0^+0^+$          & 1001(4)   & 24(16)      \\
				$\bullet$ 	& 			& $s\bar{s}$                 & 1073(10) & 0          \\ 
				$\bullet$ 	& 			& $s\bar{s}+K\bar{K}$        & 927(18)  & 1(3)       \\
				$\bullet$ 	& 			& $s\bar{s}+K\bar{K}+0^+0^+$ & 915(20)  & 2(3)       \\\hline
			\end{tabular}
		\end{normalsize}
		\caption{The results for the masses and the (half) widths for different setups of $s\bar{s}(q\bar{q})$ states in a pure RL calculation. All numerical values are given in MeV. Bullets ($\bullet$) denote whether a state is a candidate for the $f_0(980)$ and/or the $a_0(980)$.}\label{tab-a0_and_f0}
	\end{table}
	\subsection{\label{sec:4.4} Modelling Beyond Rainbow-Ladder (BRL) effects} 
	In order to get a feeling for the consequences of the BRL effects that push the pure $q\bar{q}$ ($q\in\{u,d\}$) 
	and $s\bar{s}$ mass into a physically meaningful region we introduce an additional parameter into our RL-model~(\ref{eq-RLkernel}).		
	We decrease the coupling strength in the scalar quark-antiquark BSE by a global prefactor $c<1$ (i.e. in occurrences of $K^{(2)}$ 
	in the last line of Fig.~\ref{fig-diagram2}).\footnote{
		Note that all pseudoscalar mesons such as the kaons are not affected by this procedure and therefore their
		Goldstone nature remains.}
	
	Then we performed two different test of mixing effects: first, we mix a modified $s\bar{s}$ component ($c=0.5$) with 
	an $s\bar{s}(q\bar{q})$ state and, second, we mix a modified $q\bar{q}$ component ($c=0.2$) with a $q\bar{q}(s\bar{s})$ 
	state. The first setup probes mixing of the $f_0(980)$ with strangeonium, whereas the second setup probes 
	mixing of the $a_0(980)$ and $f_0(980)$ with a conventional light-quark scalar state, as denoted in Eq.~(\ref{eq-a0_f0_components}). 
	\\[4pt]
	\textbf{Modified $\boldsymbol{s\bar{s}}$.~} 
	The results for the mixed $s\bar{s}(q\bar{q})$ state are shown in
	Tab.~\ref{tab-brl_ss} comparing RL to modified RL (mRL) calculations. We show results for pure two- and four-quark 
	states as well as the fully mixed state. We observe that the pure $s\bar{s}$ mass increases as intended by construction
	into a mass region where it can be expected to make a substantial contribution to one of the heavy scalar states,
	such as the $f_0(1500)$. The mixing effects with our light four quark state, however, are now substantially 
	smaller and the mass of the mixed state, ${M=994\:\textrm{MeV}}$, is very close to the one of the pure four-quark 
	state, ${M=1001\:\textrm{MeV}}$, which remains unchanged from table \ref{tab-a0_and_f0}. 
	
	We conclude: in this setup the mass of the mixed state is dominated by the $K\bar{K}$-component and mixing with a 
	putative conventional scalar $s\bar{s}$-meson (or such a component) is almost absent. 
	%
	\begin{table}[]
		\begin{normalsize}
			\begin{tabular}{ccccccc}\hline
				\multirow{2}{*}{$f_0$}&\multirow{2}{*}{$a_0$}&\multirow{2}{*}{setup}     &\multicolumn{2}{c}{RL}&\multicolumn{2}{c}{mRL} \\
				&                      &                           & $M$     & $\Gamma/2$ & $M$     & $\Gamma/2$   \\ \hline\hline
				$\bullet$             &                      & $s\bar{s}$                & 1073(10)& 0          & 1479(39)& 0            \\
				$\bullet$             & $\bullet$            & $K\bar{K}+0^+0^+$         & 1001(4) & 24(16)     & 1001(4) & 24(16)       \\
				$\bullet$             &                      & $s\bar{s}+K\bar{K}+0^+0^+$& 915(20) & 0          & 994(7)  & 0            \\\hline 
			\end{tabular}
		\end{normalsize}
		\caption{Comparison of results for the $s\bar{s}(q\bar{q})$ state in a pure RL calculation in comparison with a modified RL (mRL) calculation as described in the main text. The notation is the same as in table \ref{tab-a0_and_f0}.}\label{tab-brl_ss}
		\begin{normalsize}
			\begin{tabular}{ccccccc}
				\hline
				\multirow{2}{*}{$f_0$} & \multirow{2}{*}{$a_0$} & \multirow{2}{*}{setup}    & \multicolumn{2}{c}{RL} & \multicolumn{2}{c}{mRL} \\
				&                        &                           & $M$      & $\Gamma/2$  & $M$       & $\Gamma/2$  \\ \hline\hline
				$\bullet$              & $\bullet$              & $q\bar{q}$                & 661(8)   & 0           & 1288(38)  & 0           \\
				$\bullet$              & $\bullet$              & $K\bar{K}+0^+0^+$         & 1001(4)  & 24(16)      & 1001(4)   & 24(16)       \\
				$\bullet$              & $\bullet$              & $q\bar{q}+K\bar{K}+0^+0^+$& 644(6)   & 0           & 999(4)    & 17(8)        \\\hline 
			\end{tabular}
		\end{normalsize}
		\caption{Comparison of results for the $q\bar{q}(s\bar{s})$ state in a pure RL calculation in comparison with a modified RL (mRL) calculation as described in the main text. The notation is the same as in table \ref{tab-a0_and_f0}.}\label{tab-brl_qq}
	\end{table}
	\\[4pt]
	\textbf{Modified $\boldsymbol{q\bar{q}}$.~} 
	The results for the mixing of the four-quark state with a putative heavy scalar 
	$q\bar{q}$ state are shown in Tab.~\ref{tab-brl_qq}. We have chosen the extra parameter such that the mass of the (much too light)
	rainbow ladder scalar meson is lifted from 661 MeV to 1288 MeV, which is at the lower end of the mass region of the lightest 
	heavy scalar states, the $f_0(1370)$ and the $a_0(1450)$. The mass of the pure four-quark state is unchanged (1001 MeV) as
	given above in Tab.~\ref{tab-a0_and_f0}. In the RL calculation, we see that the mixed state is clearly dominated by the RL 
	$q\bar{q}$; the two states lie in the same ballpark and the $q\bar{q}s\bar{s}$ mass is far off. For the mRL calculation
	it is the other way round: within the error bars, the mixed state has the same mass and width as the pure four-quark state 
	{(1001~MeV vs. 999~MeV)}. Therefore, again, mixing between the four-quark state and the conventional heavy scalar meson state
	is negligible. Remarkably, we observe an even weaker mixing as compared to the case with the mRL $s\bar{s}$ component. 
	
	We conclude: even with the lighter $q\bar{q}$-state the mixing effects are hardly relevant. We regard this as one of the main 
	results of this work. In the present framework both, the $a_0$ and the $f_0$, are dominated by their $K\bar{K}$-component. 
	However, we wish to emphasise again that a more complete study of the nature of the $a_0$ and the $f_0$ also needs to take 
	$\pi\pi$ and $\pi \eta$ components into account, which are missing in the present study. These components will certainly compete 
	with the $K\bar{K}$-component, but we do not expect that their inclusion will change the overall dominance of meson-meson
	components as compared to the conventional $q\bar{q}$-admixtures.  
	\\[4pt]
	\textbf{Unphysical pion mass.~}
	For a meaningful comparison with lattice results such as \cite{Alexandrou:2017itd} it is also interesting to study what 
	happens when we adapt our pion masses to the (heavier) ones used in the lattice simulation. Indeed, our choice of 
	$m_q=20\:\textrm{MeV}$ for the light ($u/d$) quark mass leads to $m_\pi=328\:\textrm{MeV}$ (Tab.~\ref{tab-heavyheavy}) 
	in close proximity to the lattice value. We found that the physics does not change: the mixed state is still dominated 
	by the $K\bar{K}$ component as compared to the conventional $q\bar{q}$-admixture. As expected, the mass and width increased 
	in this process so that we arrive at the following values for the fully mixed state:
	\begin{equation}
		(M-\textrm{i}\:\Gamma/2)_{m_\pi=328\:\textrm{MeV}}=1090(3)-\textrm{i}\:55(21)\:\textrm{MeV}
	\end{equation}
    Again, the additional inclusion of the $\pi\pi$ and $\pi \eta$ components may change this value considerably. 
	\section{\label{sec:5} Summary and conclusions}\label{sec:sum}
	In this work we investigated the properties of scalar mesons in the light and strange quark sector ($q\in u,d,s$) using 
	coupled set of two-body Bethe-Salpeter equations (BSEs) that takes care of meson-meson components, diquark-anti-diquark 
	components and conventional $q\bar{q}$-components. Improving an earlier approach \cite{Heupel:2012ua}, we were furthermore
	able to study the analytic structure of the states in the complex energy plane. Our lowest mass scalar state has a complex 
	pole in the second Riemann-sheet indicating a substantial width and it is by far dominated by the $\pi\pi$ contribution; 
	admixtures from the $q\bar{q}$-components or from diquarks are insignificant. We identify this $\pi\pi$-resonance with the
	$\sigma/f_0(500)$ in (qualitative) agreement with our previous studies \cite{Heupel:2012ua,Eichmann:2015cra,Santowsky:2020pwd} 
	and other approaches like chiral effective theory \cite{Pelaez:2015qba} and lattice QCD \cite{Briceno:2016mjc}.	
	We traced this property of the $\sigma/f_0(500)$ back to the effects of dynamical chiral symmetry breaking and verified, 
	that the $\pi-\pi$-dominance slowly turns into a $q\bar{q}$-dominance by increasing the up/down-quark mass. This transition
	happens in the region of pion masses between 300-500 MeV.	
	
	Taking the strange quark into account, we also solved heavy-light mixed BSEs using candidates for the isospin partners $f_0(980)$ 
	and the $a_0(980)$. Here we focused on the interplay between a four-quark $K\bar{K}$-component, a diquark-anti-diquark-component 
	and potential admixtures from conventional $q\bar{q}$- and $s\bar{s}$-components. With masses for the conventional scalar quarkonia 
	in the region between 1250-1500 MeV we find almost no mixing effects between the conventional states and the four-quark-components.
	Furthermore, again, the four-quark state is dominated by the $K\bar{K}$-component, whereas diquark-components are irrelevant. 
	Our states acquire masses slightly above the $K\bar{K}$-threshold around 1 GeV and are therefore identified with the $f_0(980)$ 
	and the $a_0(980)$. We are, however, aware that this is not the full story. For technical reasons we have not yet included 
	potentially important $\pi\pi$ and $\pi\eta$-contributions \cite{Dudek:2016cru,Guo:2016zep} to these states; this is relegated 
	to future work.
	
	A general result of our approach to four-quark states in the light \cite{Heupel:2012ua,Eichmann:2015cra,Santowsky:2020pwd}, 
	strange and heavy-light sector \cite{Wallbott:2019dng,Wallbott:2020jzh} is the dominance of meson-meson components over 
	diquark-antidiquark components. This dominance is a dynamical issue which is based on a model-independent fact: the 
	interactions between colour-singlet antiquark-quark and colour-octet quark-quark pairs is different by a simple colour 
	factor of $1/2$. Thus, almost always the meson components of a four-quark state are lighter than the diquark components 
	and are therefore dominating\footnote{The exception are heavy-light states with open flavour, as discussed 
	in \cite{Wallbott:2020jzh}.}. This is exceptionally so in the light quark sector, since 
	chiral symmetry breaking and the associated (pseudo-)Goldstone-boson nature of pseudoscalar components enhance the 
	mass difference to the diquark components. Thus, very naturally and model-independent, light scalar four-quark states 
	are dominated by their pseudoscalar meson components.

	\subsection*{Acknowledgements}
	We are grateful to Gernot Eichmann and Marc Wagner for useful discussions. 
    This work was supported by the DFG grant FI 970/11-1, by the Helmholtz Research Academy Hesse for FAIR (HFHF) 
    and by the GSI Helmholtzzentrum f\"{u}r Schwerionenforschung.

	\begin{table*}
		\centering\normalsize
		\begin{tabular}{cccc|ccccc}
			\hline
			$a$ & \multicolumn{1}{c}{$m_q\:\text{[MeV]}$} & $m_s\:\text{[MeV]}$ & $\Lambda\:[\text{GeV}^2]$ & $m_\pi\:[\text{MeV}]$ & $f_\pi\:\text{[MeV]}$ & $m_{qq}\:[\text{MeV}]$ & $m_K\:\text{[MeV]}$ & $m_{sq}\:\text{[MeV]}$ \\ \hline\hline
			0.8 & 4.2                                      & 84.0                & 0.78                      & 140.6                 & 93.0                  & 770.7                  & 499.6               & 1051.5                 \\
			0.9 & 4.0                                      & 85.0                & 0.74                      & 139.8                 & 92.7                  & 785.2                  & 500.1               & 1072.7                 \\
			1.0 & 3.8                                      & 85.5                & 0.71                      & 139.4                 & 93.0                  & 801.3                  & 500.3               & 1108.5                 \\
			1.1 & 3.6                                      & 86.0                & 0.68                      & 138.5                 & 92.7                  & 810.4                  & 499.8               & 1126.5                 \\
			1.2 & 3.5                                      & 86.0                & 0.66                      & 140.1                 & 93.4                  & 826.4                  & 500.0               & 1146.3                 \\ \hline
		\end{tabular}
		\caption{We show different parameter sets of the Maris-Tandy model which yield physical pion and kaon properties ($m_\pi, f_\pi, m_K$). As they are important as effective ingredients of a four-quark state as well, we also show the light $qq$ and heavy-light $sq$ diquark masses. It is possible as well to vary the parameter $\eta$ between 1.6 and 2.0 independently. }\label{tab-dynamic_model}
	\end{table*}
	\begin{figure*}
		\centering
		\includegraphics[width=9.5cm]{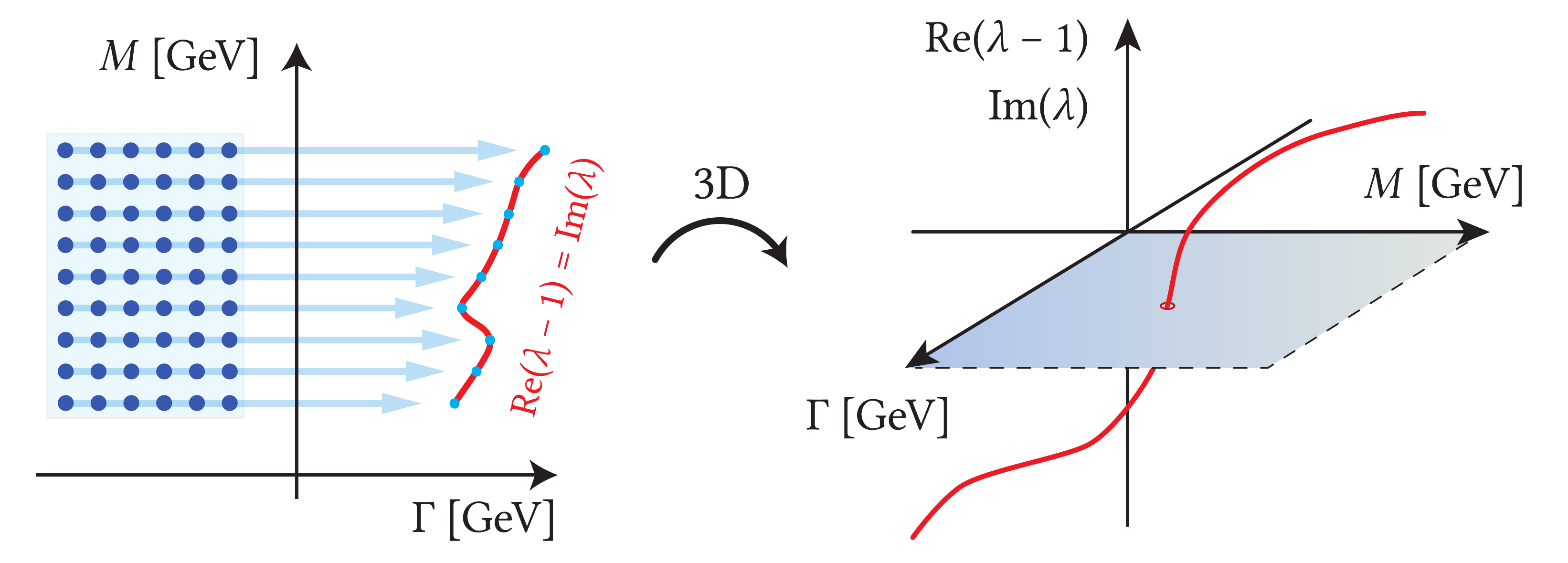}
		\includegraphics[width=8cm]{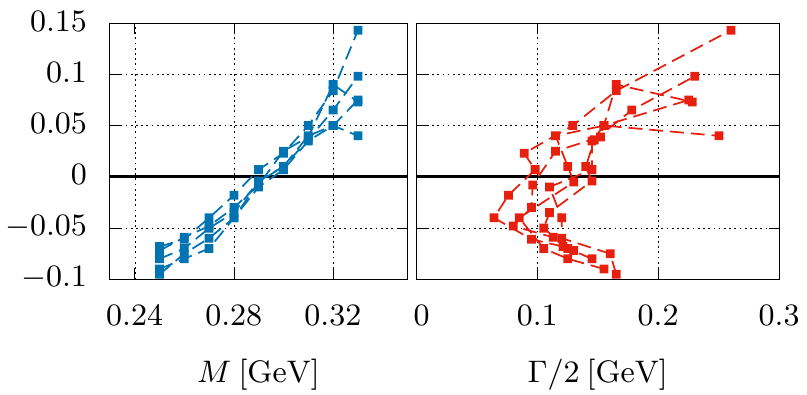}
		\caption{An illustration of a sticky curve and its computation. \textit{Leftmost panel:} Calculated input points of the eigenvalue curve (dark blue) with shadowed background. The partially transparent blue arrows and the bright blue points they point on denote one-dimensional extrapolations. The red curve is the sticky curve for which $\text{Re}(\lambda-1)=\text{Im}(\lambda)$ holds. \textit{Middle panel:} Three dimensional illustration of the sticky curve with the eigenvalue axis as the ordinate. The non-transparent surface corresponds to $\text{Re}(\lambda-1)=0=\text{Im}(\lambda)$, where the BSE is considered to have a solution, if the sticky curve crosses it. \textit{Right panels:} Projections of the sticky curves for the fully mixed $f_0(500)$ candidate in the $(M,\lambda')$ plane and in the $(\Gamma/2,\lambda')$ plane, where $\lambda'=\text{Re}(\lambda-1)=\text{Im}(\lambda)$.}\label{fig-stickycurve}
	\end{figure*}

	\newpage
\appendix
\section{Error estimates}
\subsection{Dynamic Interaction Model}\label{app:model}

	In order to solve the quark DSE and the BSEs we model our interaction kernel, see Eq.~\ref{eq-RLkernel}. The function $\alpha$ can be decomposed into an infrared and an ultraviolet part,
	\begin{align}
		\alpha(k^2) &= a\cdot\alpha_{\textrm{IR}}(k^2,\Lambda,\eta) + \alpha_{\textrm{UV}}(k^2)\,,\\
		            &= a\cdot\pi \eta^7  x^2 e^{-\eta^2 x} + \frac{2\pi\gamma_m \big(1-e^{-k^2/\Lambda_t^2}\big)}{\ln \, \left[e^2-1+\big(1+k^2/\Lambda^2_{\mathrm{QCD}}\big)^2\right]}\,,
	\end{align}
	where the UV parameters are $\Lambda_t=1$~GeV, $\Lambda_\mathrm{QCD}=0.234\,{\rm GeV}$, and $\gamma_m=12/25$ for four active quark flavours.
	While the ultraviolet part is determined by perturbation theory to ensure the logarithmic behaviour of the coupling in the far-UV, the infrared part with $x=k^2/\Lambda^2$ is the actual model that includes two parameters, $\Lambda$ and $\eta=1.8$. 
	We also attached a prefactor $a$ in front of the infrared part that scales the coupling strength linearly; the original model is restored by setting $a=1$. These parameters along with the renormalized quark masses $m_q$ and $m_s$ are chosen in a way that the physics of the pseudoscalar mesons, the pion mass $m_\pi$ and the kaon mass $m_K$, are obtained in agreement
	with the experimental values. For the pion, we also ensure that the decay constant $f_\pi$ retains its correct value. 
	In order to estimate a model error, we vary the complete set of parameters $\{a,\Lambda,m_q,m_s\}$. The parameter sets 
	and the corresponding results for the light quarks $u/d$ and $s$ are shown in Tab.~\ref{tab-dynamic_model}. 
	The parameter $\eta$ could be varied independently within the band $[1.6,2.0]$. As we wished, the pion and 
	kaon mass along with the pion decay constant are consistently physical, whereas the diquark masses are dynamic, 
	which is alright as they are not physical anyhow. All this provides a model-internal error estimation which 
	is a part of the total errors given in the results section of this work. Besides that, additional errors 
	come from extrapolations, which is described in the following section.

	\subsection{Extrapolations}\label{app:extra}
	Besides model-internal parameter variations, the extrapolations of the eigenvalue curves into the second Riemann sheet cause an error. For the extrapolations in the complex plane, we compute so-called \textit{sticky curves}. The equation is considered to be solved if the eigenvalue is one, $\lambda=1$, i.e.
	\begin{equation}
		\text{Re}(\lambda-1)=0=\text{Im}(\lambda).
	\end{equation}
	We therefore extract the (reduced) real and imaginary part in the first Riemann sheet where the eigenvalue curve is directly accessible and extrapolate to the point where they have the same value, $\text{Re}(\lambda-1)=\text{Im}(\lambda)$, thus `stick together'. For resonant states, these sticky curves cross the zero in the second Riemann sheet, or for positive $M$ and $\Gamma$, respectively. This procedure along with the curves for the fully mixed $f_0(500)$ candidate is shown in Fig.~\ref{fig-stickycurve}. As one could see, we obtain a mass around 290 MeV with a comparably small model error and a width around 120\:MeV with a larger error. The curves for other states such as the candidates for the $f_0/a_0(980)$ look very similar and are not shown here.
	\\
	
	\bibliographystyle{apsrev4-1}
	\bibliography{complex_lights}
	
\end{document}